\documentclass[prl,superscriptaddress,amsfonts,twocolumn,showpacs,floatfix]{revtex4-1}
\usepackage{amsmath}
\usepackage{amsfonts}
\usepackage{amssymb}
\usepackage{bm}
\usepackage{txfonts}
\usepackage{tabularx}
\usepackage[dvips]{graphicx,color}

\newcommand{\pco}{PdCoO$_2$}

\newcommand{\subm}[1]{_\mathrm{#1}}
\newcommand{\vF}{\bm{v}\subm{F}}
\renewcommand{\deg}{^\circ}

\begin{document}
\title{
Extremely Large Magnetoresistance in the Nonmagnetic Metal PdCoO$_{2}$
}

\author{Hiroshi Takatsu}
\affiliation{Department of Physics, Tokyo Metropolitan University, Tokyo 192-0397, Japan}
\affiliation{Department of Physics, Graduate School of Science, Kyoto University, Kyoto 606-8502, Japan}

\author{Jun J. Ishikawa}
\affiliation{Department of Physics, Graduate School of Science, Kyoto University, Kyoto 606-8502, Japan}
\affiliation{Institute for Solid State Physics, University of Tokyo, Kashiwa 277-8581, Japan}

\author{Shingo Yonezawa}
\affiliation{Department of Physics, Graduate School of Science, Kyoto University, Kyoto 606-8502, Japan}

\author{Harukazu Yoshino}
\affiliation{Department of Material Science, Graduate School of Science, Osaka City University, Osaka 558-8585, Japan}

\author{Tatsuya Shishidou}
\affiliation{Department of Quantum Matter, ADSM, Hiroshima University, Higashihiroshima, Hiroshima 739-8530, Japan}

\author{Tamio Oguchi}
\affiliation{Institute of Scientific and Industrial Research, Osaka University, Ibaraki, Osaka 567-0047, Japan}

\author{Keizo Murata}
\affiliation{Department of Material Science, Graduate School of Science, Osaka City University, Osaka 558-8585, Japan}

\author{Yoshiteru Maeno}
\affiliation{Department of Physics, Graduate School of Science, Kyoto University, Kyoto 606-8502, Japan}
\date{\today}

\begin{abstract}
Extremely large magnetoresistance is realized in the nonmagnetic layered metal PdCoO$_2$.
In spite of a highly conducting metallic behavior with a simple quasi-two-dimensional hexagonal Fermi surface,
the interlayer resistance reaches up to 35000\% for the field along the $[1\bar{1}0]$ direction.
Furthermore,
the temperature dependence of the resistance becomes nonmetallic for this field direction, 
while it remains metallic for fields along the [110] direction.
Such severe and anisotropic destruction of the interlayer coherence by a magnetic field on a simple Fermi surface 
is ascribable to orbital motion of carriers on the Fermi surface driven by the Lorentz force, but seems to have been largely overlooked until now.
\end{abstract}

\maketitle

Finding new systems exhibiting a large resistance change by a magnetic field
has driven crucial progress in both condensed matter physics and device application.
The most well-known examples are 
the giant magnetoresistance in magnetic multilayers~\cite{BaibichPRL1988,BinaschPRB1989} 
and
colossal magnetoresistance in manganites~\cite{A.D.Torres,Y.Tokura}, 
both of which rely on coupling between spin configuration and charge transport.
Even among nonmagnetic materials,
the magnetoresistance (MR) may become large in systems 
such as semimetals with Fermi surface (FS) compensation~\cite{YangScience1999,X.DuPRL2005,Kasahara2007.PhysRevLett.99.116402}. 
Here we report extremely large MR and metal-nonmetal crossover in a highly conducting nonmagnetic metal PdCoO$_2$.
Realized with a simple quasi-two-dimensional (quasi-2D) FS, 
this huge MR simply originates from the orbital motion of the highly conducting electrons.

The delafossite compound PdCoO$_2$ has a layered hexagonal structure with the space group $R\bar{3}m$
consisting of alternating stacking of Pd triangular layers and CoO$_2$ triangular slabs~\cite{Shannon1971,Tanaka1998,Itoh1999,Seshadri1998,Takatsu2007,V.EyertChmMater2008,H.J.NohPRL2009,K.Kim2009,K.P.OngPRL2010,K.P.OngPRB2010}.
The metallicity is predominantly attributed to the Pd $4d$ electrons~\cite{Higuchi1998, Hasegawa2001,H.J.NohPRL2009};
the densities of states of Co and O, as well as Pd $5s$, are very low at the Fermi level~\cite{K.P.OngPRB2010}.
The band calculations indicate that 
the FS consists of a single rounded hexagonal prism~\cite{Seshadri1998,V.EyertChmMater2008,K.Kim2009,K.P.OngPRL2010,K.P.OngPRB2010,Hicks2012.PhysRevLett.109.116401}.
The electronic specific-heat coefficient is $\gamma\simeq1$~mJ/mol K$^2$,
indicating that electron correlation in PdCoO$_2$ is not strong.
The carrier density $n$ estimated from the Hall coefficient is $n_{\mathrm{obs}}=1.6\times10^{22}$~cm$^{-3}$, 
consistent with one electron carrier per formula unit~\cite{Takatsu2007,H.TakatsuPRL2010}.
This agreement indicates that this oxide is a high-carrier-density, free-electron-like system.
Reflecting the 3$d^6$ low-spin state of Co$^{3+}$ with zero total spin,
PdCoO$_2$ is nonmagnetic in the whole 
temperature and field ranges investigated~\cite{Tanaka1996,Itoh1999,Takatsu2007,H.J.NohPRB2009}.

In this Letter,
we report that the electrical resistivity along the $c$ axis $\rho_c$ 
is surprisingly enhanced by 
the application of the in-plane magnetic field along the $[1\bar{1}0]$ direction,
reaching 35000\% of the zero-field resistance at 2~K and 14~T,
and continues to increase linearly with field.
We also found that the temperature dependence of $\rho_c$ for this field direction 
exhibits a metal-nonmetal crossover at around 120~K.
This behavior is sensitively suppressed by a small tilt or rotation of the magnetic field.
Semiclassical calculations of the MR  
based on the tight-binding band structure of PdCoO$_2$ 
qualitatively reproduce the observed field-angle dependence of the MR.
Thus, this extremely large MR is a generic but overlooked phenomenon in a highly conducting
system with a simple quasi-2D FS; it is due to the destruction of interlayer coherence 
by an orbital motion of the carriers on the FS driven by the Lorentz force.

\begin{figure}[b]
\begin{center}
 \includegraphics[width=0.4\textwidth]{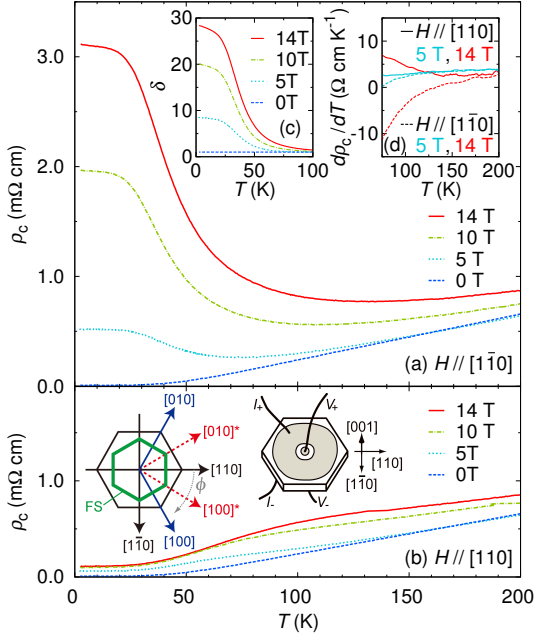}
 \caption{
(color online) 
Temperature dependence of the interlayer resistivity
$\rho_{c}$ for several magnetic fields along 
the (a) $[1\bar{1}0]$ and (b) $[110]$ directions.
The strong enhancement with a nonmetallic upturn was observed 
for $H\parallel[1\bar{1}0]$.
In contrast,
the enhancement for $H\parallel[110]$ is an order 
of magnitude smaller.
The left-hand inset of (b) is a schematic drawing of the FS with its relation to 
the crystal axes and reciprocal lattice vectors.
The right-hand inset shows the wire arrangement for the measurements of $\rho_{c}$.
(c) In-plane field anisotropy ratio $\delta = \rho_c(H\parallel[1\bar{1}0])/\rho_c(H\parallel[110])$
for several field strengths.
(d) Temperature derivative of resistivities for field
along the $[1\bar{1}0]$ and $[110]$ directions at 5~T and 14~T.
}
\label{fig.1}
\end{center}
\end{figure}

Single crystals of PdCoO$_2$
used in this study are grown by a stoichiometric self-flux method.
They were characterized by the powder x-ray diffraction 
and the energy dispersive x-ray analysis~\cite{Takatsu2007}.
The out-of-plane resistivity $\rho_{c}$ was measured 
on single-domain crystals with the hexagonal plate shape
having an area of about 5~mm$^2$ and a thickness of about $0.03$~mm.
A conventional four-probe method was employed from 2 to 300~K
in a field up to 14~T
with a commercial apparatus (Quantum Design, model PPMS).
We attached two gold wires ($\phi=25~\mu{\rm{m}}$) for current and voltage ($I_+$ and $V_+$)
to one $ab$ surface of the crystal and another two wires ($I_-$ and $V_-$) to the other side. 
The wires were attached with silver paste (Dupont, 6838). 
The silver paste for the current electrodes was made into a ring shape to cover as 
wide area of the crystal surfaces as possible, 
while the voltage leads were placed at the center of the ring of the current electrodes, 
as shown in the inset of Fig.~\ref{fig.1}(b).
This wire arrangement is justified for anisotropic metals, and is indeed often used for layered compounds~\cite{Rogers1971,MaenoNature1994}.
The residual resistivity ratio,
$\rho_{c}(300~\mathrm{K})/\rho_{c}(2~\mathrm{K})$,
in zero field
is about 120 and guarantees a high sample quality.
The sample crystal was placed with its basal $ab$ plane parallel to 
the rotating stage of a single axis rotator,
which controls the azimuthal orientations of the magnetic field.
To check the possibility of a magnetic phase transition induced by 
a magnetic field,
the specific heat ($C_P$) and dc magnetic susceptibility ($M/H$) were measured
with a commercial calorimeter and with a SQUID magnetometer on 
a group of aligned crystals.

Figure {\ref{fig.1}} presents the temperature dependence 
of $\rho_{c}$ for fields along the $[1\bar{1}0]$ and $[110]$ directions.
With magnetic fields in the $ab$ plane,
$\rho_{c}$ exhibits substantial enhancement on cooling. 
In particular,
for fields along the $[1\bar{1}0]$ direction,
$\rho_{c}$ even turns into nonmetallic, exhibiting huge enhancement.
It reaches 350 times larger than the zero-field 
resistivity value at 2~K in 14~T.
This enhancement is comparable to those of 
giant magnetoresistance and colossal magnetoresistance materials~\cite{A.D.Torres,Y.Tokura}.
However, in sharp contrast with these materials,
PdCoO$_2$ exhibits a positive field response.
Moreover, as confirmed by in-field measurements of $C_P$ and $M/H$,
we did not observe any anomalies suggesting magnetic or charge order
at temperatures 
where the enhancement of $\rho_c$ develops.

When rotating the field from the $[1\bar{1}0]$ to $[110]$ directions,
$\rho_c$ sharply drops and switches to metallic,
although it slightly deviates from a typical $T$--linear dependence 
in zero field.
A small hump structure exists at temperatures
where $\rho_{c} (H\parallel[1\bar{1}0])$ starts to increase.
As is clear from Figs.~\ref{fig.1}(c) and (d),
the in-plane field anisotropy emerges at a temperature which depends on the field strength.
In fact,
$d\rho_c/dT$ for $H\parallel[1\bar{1}0]$ and $H\parallel[110]$ starts to deviate from each other
at about 100~K in 5~T, and about 180~K in 14~T.
The anisotropy ratio, $\delta \equiv \rho_c(H\parallel [1\bar{1}0]) / \rho_c(H\parallel [110])$, 
is up to 29 (2900\%) at 2~K in 14~T.
{\color [rgb] {0,0,0} This value is larger than the field anisotropy of ordinary 
nonmagnetic metals with simple sphere FSs, such as alkali metals ($<1$\%)~\cite{A.Overhauser}.
It is also noteworthy 
that this anisotropy value is even larger than a typical ratio of 
the anisotropic magnetoresistance of magnetic metals ($<$2\%--3\%)~\cite{I.A.Campbell}.}

\begin{figure}[b]
\begin{center}
 \includegraphics[width=0.40\textwidth]{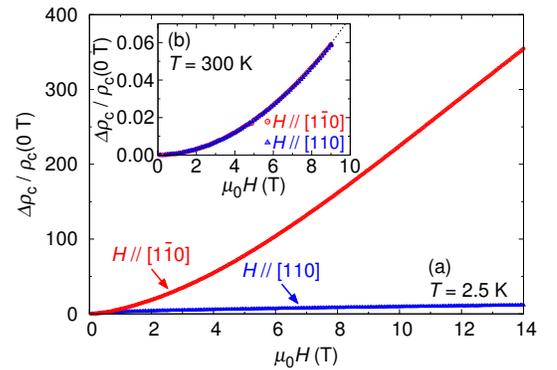}
 \caption{
(color online) 
Field dependence of the magnetoresistance $\varDelta\rho_{c}/\rho_{c}(H=0)$
at (a)  2.5~K and (b) 300~K.
The magnetoresistance at low temperature is much greater 
for field along the $[1\bar{1}0]$ direction.
At 300 K, 
the magnetoresistance is nearly independent of the in-plane
field direction and exhibits an $H^2$ dependence as indicated with the dotted curves.
}
\label{fig.2}
\end{center}
\end{figure}

Figure~\ref{fig.2} shows the field dependence of the MR ratio,
$\varDelta \rho_{c}(H)/\rho_{c}(H=0)= \rho_{c}(H)/\rho_{c}(H=0)-1$, 
for the $[1\bar{1}0]$ and $[110]$ field directions.
As observed in the temperature dependence,
$\varDelta \rho_{c}(H)/\rho_{c}(H=0)$ for the $[1\bar{1}0]$ field direction
exhibits steep increase with increasing field strength, 
reaching up to 350 at 2.5~K.
$\varDelta \rho_{c}(H)/\rho_{c}(H=0)$ for the $[110]$ field direction 
also increases but it is an order of magnitude smaller 
than that for the field along $[1\bar{1}0]$.
We note that even at 300~K,
the magnetoresistance amounts to 6\% in 9~T
and shows the isotropic in-plane field dependence [Fig.~\ref{fig.2}(b)].
The samples used for Figs.~\ref{fig.2}(a) and (b) are different, 
but we confirmed that both samples show essentially 
the same behavior.
At lower in-plane fields, at 2.5~K,
$\rho_c$ exhibits isotropic field response.
It follows the $H^{2}$ dependence 
expected from the orbital motion of conduction electrons~\cite{Ziman2}.
However, in fields above 0.5~T,
a clear anisotropy emerges with different field dependence;
$\rho_{c}(H\parallel[1\bar{1}0])$ varies as $H^{1.5}$, while $\rho_{c}(H\parallel[110])$ is 
proportional to $H^{0.5}$.
This fact implies that $\omega_c\tau$,
where $\omega_c$ is the cyclotron frequency and
$\tau$ is the average relaxation time of charge carriers,
becomes large in fields above 0.5~T (at 2.5~K)
and the field response of the carrier mobility 
goes into an intermediate field region.
In such a region, 
several orbital motions may contribute to 
the field dependence of the resistivity,
leading to the super- or sublinear field dependence~\cite{Ziman2}.
{\color [rgb] {0,0,0} As another aspect of the field dependence,
it is known that such a $H^{1.5}$ dependence can originate from 
the out-of-plane incoherent transport~\cite{HusseyPRB1998},
in which a large number of in-plane scatterings of conduction electrons 
occur before electrons hop or tunnel to a neighboring plane.}

\begin{figure}[b]
\begin{center}
 \includegraphics[width=0.40\textwidth]{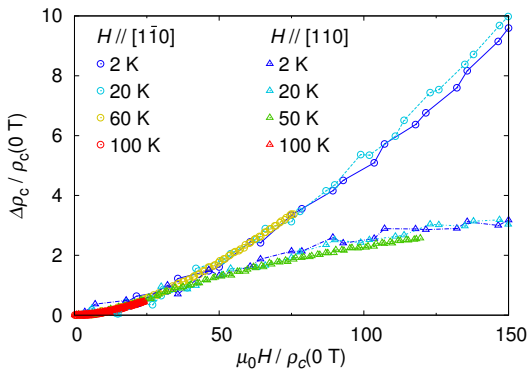}
 \caption{
(color online) 
The so-called Kohler plot:
the magnetoresistance ratio versus field divided by 
the zero-field resistivity.
}
\label{fig.3}
\end{center}
\end{figure}
Figure 3 is the so-called Kohler plot, in which 
$\varDelta \rho_{c}(H)/\rho_{c}(H=0)$ is plotted against 
$\mu_0H/\rho_c(H=0)$.
The universality among different temperatures is satisfied for each field 
direction;
this indicates that the scattering process is well explained by
a single relaxation rate $\tau$ and the dominant scattering process is not changed by field and temperature~\cite{Ziman2}.

\begin{figure}[b]
\begin{center}
 \includegraphics[width=0.4\textwidth]{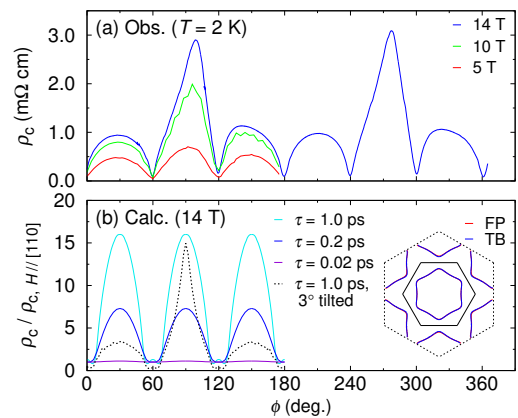}
 \caption{
(color online) 
(a) Observed azimuthal field angle $\phi$ dependence of the interlayer resistivity $\rho_{c}$ of PdCoO$_2$ at 2~K, for $\mu_0H = 5$, 10, and 14~T.
(b) Calculated $\phi$ dependence of $\rho_{c}$ at $\mu_0H = 14$~T and 
$\tau=1.0$, 0.2, and 0.02~ps, normalized by $\rho_c$ for $H\parallel [110]$. 
We also present $\rho_c$ at 1.0~ps for fields slightly tilted from 
the $ab$ plane with the broken curve. 
For this calculation, the field-rotation plane is assumed to be 
tilted by $3\deg$ from the $[1\bar{1}0]$ direction in the $ab$ plane
to the $[001]$ direction.
The inset of (b) presents the Fermi surface cross section at $k_z=0$ for the first-principles (FP) band calculation and the tight-binding (TB) model.
The solid hexagon indicates the first Brillouin zone (BZ) for the hexagonal representation and the broken hexagon the first BZ for the rhombohedral representation.
The latter is shown to present a larger portion of the Fermi surface.
}
\label{fig.4}
\end{center}
\end{figure}

Figure \ref{fig.4}(a) represents the in-plane field-angle $\phi$ dependence of $\rho_c$ at 2~K 
for several field strengths. 
The periodicity of $\rho_c$ is essentially $60\deg$ at and below 14~T. 
The observed asymmetry is attributed to a small misalignment of 
the magnetic field with respect to the $ab$ plane. 
In fact, we have confirmed that $\rho_c$ for high fields dramatically changes even with 
a few-degree misalignment of the magnetic field away from the $ab$ plane~\cite{note_misalignment_of_sample}. 
As we will explain later, the misalignment effect 
is also reproduced by the calculation [Fig.~\ref{fig.4}(b)].
The observed $\rho_c$ oscillation is consistent with the field-strength dependence: 
$\rho_c$ exhibits minima for $\bm{H}\parallel [110]$ and maxima for $\bm{H}\parallel [1\bar{1}0]$.

In order to resolve the origin of the anomalously large MR effect in a simple metal, 
we have calculated the MR of PdCoO$_2$ by solving the semiclassical Boltzmann equation:
\begin{align}
\sigma_{ij}(\bm{B}) &= \frac{2e^2}{V}\sum_{\bm{k}}\left(-\frac{\mathrm{d}f_0(\varepsilon)}{\mathrm{d}\varepsilon}\right)v_i(\bm{k}(0))\int^0_{-\infty}v_j(\bm{k}(t))e^{t/\tau}\mathrm{d}t,\label{eq:Boltzmann}
\end{align}
where $\sigma$ is the conductivity tensor, 
$e$ is the elementary charge, $V$ is the volume of the sample, 
$f_0(\varepsilon)$ is the Fermi distribution function at $T=0$, 
$v_i$ is the Fermi velocity along the direction $i = x, y, z$, and $\tau$ is the relaxation time.
Then the resistivity tensor $\rho$ is obtained as the inverse of the conductivity tensor: $\rho=\sigma^{-1}$.
Note that the assumption of a single relaxation time is justified by the fact that Kohler's rule is satisfied in the present field range.
The orbital motion of conduction electrons in the magnetic field causes the time evolution of $\bm{k}(t)$, which is expressed as 
\begin{align}
\frac{\mathrm{d}\bm{k}(t)}{\mathrm{d}t} = -\frac{e}{\hbar}\vF\times\bm{B} = -\frac{e}{\hbar^2}\bm{\nabla}_{\bm{k}}\varepsilon\times\bm{B}.
\end{align}
The solution \eqref{eq:Boltzmann} is numerically calculated with $N\times N\times N$ ($N = 2^5$ or $2^6$) meshes 
in the reciprocal space.
This process is essentially identical to that used for the quasi-one-dimensional conductor (TMTSF)$_{2}X$ in Ref.~\cite{Yoshino1999.JPhysSocJpn.68.3027}.

We attempted to calculate the MR using a model dispersion relation $\varepsilon(\bm{k})$ that
approximately reproduces the Fermi surface obtained from the first-principles calculations:
\begin{align*}
\varepsilon(\bm{k}) = &-2t_1[\cos(\bm{k}\cdot\bm{a}) + \cos(\bm{k}\cdot\bm{b}) + \cos(-\bm{k}\cdot(\bm{a} + \bm{b}))]\nonumber\\
&-2t_2\cos(\bm{k}\cdot\bm{c})\nonumber \\
&- 2t_3[\cos^2(\bm{k}\cdot\bm{a}) + \cos^2(\bm{k}\cdot\bm{b}) + \cos^2(-\bm{k}\cdot(\bm{a} + \bm{b}))],
\label{eq:TB-dispersion}
\end{align*}
where $t_i$ are tight-binding-like phenomenological hopping energies, $\bm{a}$, $\bm{b}$, and $\bm{c}$ are crystalline unit vectors of the hexagonal representation of the $R\bar{3}m$ structure of \pco.
We chose $t_1\sim 1.0$~eV, $t_2\sim 0.01$~eV, and $t_3\sim 0.14$~eV, so that the resulting FS matches that obtained from the first-principles band calculation~\cite{Shishidou2} as shown in the inset of Fig.~\ref{fig.4}(b).
Note that $t_1$ is the in-plane nearest-neighbor (i.e.~Pd-Pd) hopping, $t_2$ is the interlayer hopping energy for the $c$-axis unit cell length (i.e.~three Pd layers), and $t_3$ is the in-plane third-nearest-neighbor (i.e.~Pd-Pd-Pd) hopping.
Thus, $t_1$, $t_2$ and $t_3$ should have approximate relations to the in-plane nearest-neighbor hopping $t\subm{NN}$, the in-plane next-nearest neighbor hopping $t\subm{NNN}$, and the $c$-axis nearest neighbor hopping $t_{zz}$ used in the recent de Haas-van Alphen study~\cite{Hicks2012.PhysRevLett.109.116401} as $t_1 = t\subm{NN}$, $t_2\sim t_{zz}/3$, and $t_3\sim-\sqrt{3}t\subm{NNN}/2$.
This is indeed the case since $t\subm{NN}=1$~eV, $t\subm{NNN}=-0.23$~eV, and $t_{zz} = 0.042$~eV is deduced from the de Haas-van Alphen results~\cite{Hicks2012.PhysRevLett.109.116401.SuppleMat}.
{\color [rgb] {0,0,0} 
We comment here that effects of the spin-orbit interaction to the MR 
are probably negligible.
This is because the spin-orbit interaction  band splitting around the Fermi level is negligibly small 
although the coupling constant of the spin-orbit interaction, $\zeta$, 
is relatively large at the Pd site ($\zeta \sim 0.22$~eV)
from our band calculation.
The absence of the band splitting around the Fermi level is attributed to the fact that the FS 
consists of one band of Pd $d$ electrons, being well separated in energy from the other bands~\cite{Shishidou2}.}

\begin{figure}[b]
\begin{center}
 \includegraphics[width=0.4\textwidth]{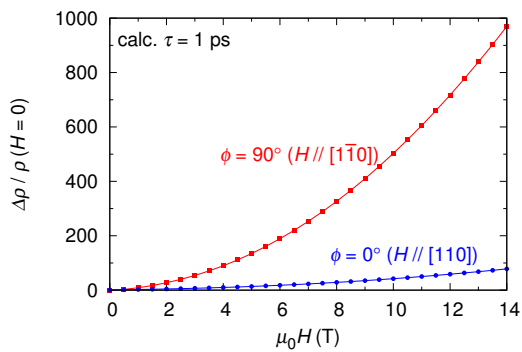}
 \caption{
(color online) 
Calculated field dependence of the out-of-plane magnetoresistance $\varDelta\rho_c/\rho_c(H=0)$ for $\tau = 1$~ps. 
The solid lines just connect the discrete calculated data points.
The observed large anisotropy in $\varDelta\rho_c/\rho_c(H=0)$ is well reproduced.}
\label{fig.5}
\end{center}
\end{figure}

Figure \ref{fig.4}(b) represents the calculated $\rho_c=\rho_{zz}$ for $\mu_0H=14$~T and $\tau=1.0$, 0.2, and 0.02~ps. 
The calculation qualitatively reproduces the observed behavior: i.e., minima in $\rho_{zz}$ for $\bm{H}\parallel [110]$.
We should also point out that the in-plane MR anisotropy grows rapidly as $\tau$ becomes longer:
$\rho(0\deg)/\rho(90\deg)$ reaches 15 for $\tau=1$~ps.
We also calculated the field dependence of $\varDelta\rho_c/\rho_c(H=0)$ for $\tau=1$~ps as plotted in Fig.~\ref{fig.5}.
This calculation also captures the enhancement of $\rho_c$ by several hundred times for $\bm{H}\parallel [1\bar{1}0]$ (i.e. $\phi=90^\circ$).
These agreements strongly indicate that the observed large MR is closely related to the band structure near the Fermi energy 
and essentially ascribable to the orbital motions of conduction electrons.
In particular, the in-plane hexagonal anisotropy of the FS 
(i.e., the $t_3$ term in the dispersion or the $k_{31}$ and $k_{60}$ terms in Ref.~\cite{Hicks2012.PhysRevLett.109.116401}) plays an important role in the anisotropy in the MR:
For $\bm{H}\parallel [110]$, two planes of the FS are nearly perpendicular to the magnetic field.
Since $\vF\propto -\nabla_{\bm{k}}\varepsilon(\bm{k}\subm{F})$ is perpendicular to the FS, the Lorentz force $\bm{F}\subm{L}\propto \vF\times \bm{B}$ is nearly zero for carriers on this part of the FS.
Thus, roughly speaking, $2/6$ of the carriers do not contribute to the enhancement of the MR.
In contrast, for $\bm{H}\parallel [1\bar{1}0]$, almost all conduction electrons on the FS feel the Lorentz force and contribute to the MR enhancement. 
In particular, for the carriers on the two FS planes parallel to the magnetic field, $|\bm{F}\subm{L}|$ is maximized resulting in the large MR enhancement.
Similar cases have been found in quasi-1D or -2D organic conductors~\cite{Osada1996,Yoshino1999.JPhysSocJpn.68.3027,Kartsovnik2006.PhysRevLett.96.166601}.
In addition, closed orbits due to the FS warping along the $k_z$ direction may also contribute to the MR. 
Indeed, the MR for fields nearly parallel to the $[1\bar{1}0]$ direction is quite sensitive to the field alignment along the conducting $ab$ plane: even a tiny misalignment of the field out of the conducting plane reduces the MR [see Fig.~\ref{fig.4}(a)].
The misalignment effect is also reproduced by the calculation, as shown with 
the broken curve in Fig.~\ref{fig.4}(b).
According to the calculation, even $3\deg$ field misalignment 
results in a reduction of MR by 75\%.

The calculation reveals that a long $\tau$ is essential for the MR of \pco.
Indeed, it has been pointed out that the scattering rate ($\sim 1/\tau$) of \pco\ is surprisingly small~\cite{Hicks2012.PhysRevLett.109.116401}.
We can estimate the actual $\tau$ using the observed in-plane mean free path $l=\langle\vF^2\rangle^{1/2} \tau \sim 20~\mu$m from Refs.~\cite{Hicks2012.PhysRevLett.109.116401} and 
\cite{Hicks2012.PhysRevLett.109.116401.SuppleMat}, and the mean Fermi velocity $\langle\vF^2\rangle^{1/2} \sim 1.5\times 10^6$~m/s from our calculation: we obtain $\tau\sim 10$~ps, which is even larger than our calculation limit~\cite{comment_tau}.
Thus, the unusual suppression of scattering is the key ingredient of the observed large MR.
Note that the extremely large MR 
in a simple metal has been overlooked for many years.
The reason is probably that a layered material with such a high mobility has been quite rare so far.

To summarize, we have discovered the {\color [rgb] {0,0,0} extremely large} 
magnetoresistance  
reaching $\varDelta\rho(H=14~\mathrm{T})/\rho(H=0~\mathrm{T})= 35000$\% in the quasi-two-dimensional, nonmagnetic metal PdCoO$_2$.
This MR is surprising since its electronic structure is very simple.
Based on the semiclassical calculation, we demonstrate that the observed MR is
closely related to the Lorentz-force-driven orbital motion of the high-mobility charge carriers.
The present finding marks PdCoO$_2$ as the first single-band simple metal exhibiting extremely large MR 
and furthermore opens a route to apply this oxide to industrial devices such as magnetic sensors.

We would like to thank 
N. Hussey, X. Xiaofeng, M. Kriener, S. Kittaka, 
C. Michioka, and K. Yoshimura for experimental advice and supports.
We also acknowledge A. Mackenzie, C. Hicks, 
K. Ishida, Y. Ihara, Y. Nakai,
H. Kadowaki, and R. Higashinaka for useful discussions.
This work was supported by the MEXT 
Grants-in-Aid for Scientific Research 21340100,
for Research Activity Start-up 22840036, for Young Scientists (B) 24740240
and for the Global COE program ``The Next Generation of Physics, Spun from Universality and Emergence.''

{\it Note added in proof} -- We recently became aware of the
work on PtSn$_4$ \cite{MunPRB2012} reporting a magnetoresistance even
greater than that which we report here. The Fermi surface
of PtSn$_4$ is rather complicated and the mechanism of the
large MR is attributed to the compensation of mobilities of
different carriers. We thank P. Canfield for directing our
attention to this work.

\bibliography{reference}

\begin{thebibliography}{37}
\bibitem [{not()}]{note_misalignment_of_sample}%
  \BibitemOpen
  \href@noop {} {}\bibinfo {note} {From the polar angle $\theta$ dependence of
  $\rho_c$ between the $[001]$ axis and the $[1\bar{1}0]$ axis, the
  misalignment of the angle in $\rho_c (\phi)$ was roughly estimated to be
  $\theta\simeq3^\circ$, assuming that the azimuthal angle $\phi$ was just in
  the $[1\bar{1}0]$ direction. For this estimation, we measured another sample
  with the single axis rotator.}\BibitemShut {Stop}%
\end{thebibliography}
\end{document}